\documentstyle[11pt,a4]{article}
\def\largelinestretch{\renewcommand{\baselinestretch}{1.5}}
\largelinestretch\small\normalsize
\def\largelinestretch{\renewcommand{\baselinestretch}{1.1}}
\title{\Large
What can we learn from new measurements of Dalitz plot
parameters for K $\rightarrow 3 \pi$ decays ? }
 \author{
A.A.Bel'kov${}^1$,
G.Bohm${}^2$,
F.Matth\"ai${}^3$,
A.V.Lanyov${}^1$,
A.Schaale${}^2$
\\
\\
\small
${}^1$
        Particle Physics Laboratory, Joint Institute for Nuclear Research,
\hfill\\
\small
        Head Post Office, P.O. BOX 79, 101000 Moscow, Russia
\hfill\\
\small
${}^2$
        DESY-Institute for High Energy Physics,
        Platanenallee 6, 15735 Zeuthen, Germany
\hfill\\
\small
${}^3$
        Friedrich Schiller Universit\"at, 07743 Jena,
        Physikalische Astronomische Fakult\"at
\hfill\\
\small
        Max Wien Platz 1
\hfill\\
}
\begin{document}
\largelinestretch\normalsize
   \thispagestyle{empty}
   \begin{titlepage}
   \maketitle
   \begin{abstract}
We give a simple expression in linear and quadratic Dalitz--plot
slopes which does not depend on the charge combination of the
$3\pi$ state $(K^\pm \to \pi^{\pm}\pi^{+}\pi^{-}$ or
$\pi^{\pm}\pi^{0}\pi^{0}$ and $K_L^{0} \to \pi^{+}\pi^{-}\pi^{0}$ or
$\pi^{0}\pi^{0}\pi^{0})$, if all phases between final
states are
negligible. After investigating the influence of radiative corrections,
 it is shown how new measurements, especially of
quadratic slopes in the $\pi^{\pm} \pi^{0} \pi^{0}$ channel, could help to
test theoretical predictions more stringently.\\
   \end{abstract}
   \end{titlepage}

%
\section*{Introduction}
%

At least until the advent of decisive data on CP--violation from the
B sector, the K sector remains our main source of
experimental information on this old and important question. Further
progress depends, however, not only  on very high statistics
experiments, but also on well tested effective theoretical models for
handling long--range strong interactions in a quantitative way.

As in the 3--generation quark model CP--violation is thought to be due
to a phase factor in the Cabbibo--Kobayashi--Maskawa--matrix, any
measurable first order effect (,,direct''\,  CP--violation) is
expected to disappear if certain phase shifts from (strong) final state
interactions (f.s.i.) between different components of transition amplitudes are
vanishing. In an attempt to estimate the size of CP--violation effects in
K$ \rightarrow 3 \pi$  decays \cite{papers} it was found necessary not
only to fit a number of coefficients determining the weak effective
Lagrangian from experimental data, but also to rely  on loop
calculations in deriving the f.s.i. phases. A further problem is
that the strong interaction effects in K$
\rightarrow 3 \pi$  decays {} cannot be handled
(like in K$ \rightarrow 2\pi$)
by just multiplying
final (strong) eigenstates with constant phase factors; imaginary
parts from pion loops depend on dynamical variables, and the real parts
are modified in turn by various loop contributions, among them
loops with inner K lines. Therefore the denotation as f.s.i.
phases is not quite correct. It would be
desirable to get more direct experimental information on these phases
from existing experimental data on K decays, but this turns out to be
very difficult, as we will explain below.

A well documented joint fit of isotopic K$ \rightarrow 2 \pi, 3 \pi$
amplitudes has been given by Devlin and Dickey\cite{devlin}, and  more
recent work is published by Kambor et al.\cite{kambor}. Both groups of
authors found it necessary to make assumptions on the
f.s.i. phases in $3 \pi $ states, namely that they are zero, resp.
$< 15^o$. While  general unitarity arguments lead us to agree with
these expectations concerning the constant
parts of the amplitudes, the situation may be
more complex for the coefficients ($b_{IJ}, c_{IJ}, d_{IJ}$, see below) of the
non constant parts, which vanish at the center of the Dalitz
plot.

Instead of repeating once more these fits, varying merely the data
sets used and the parameter sets to be fitted, it may be more useful
to investigate by simpler means which data are particularly important for our
question, and to check their internal consistency and the appropriateness
of the radiative corrections applied
before the data can be used to fit isospin amplitudes.
Their effect is found to be very important for our investigation, and the
standard procedure for handling them (see \cite{pdg}) may not be sufficient.

In the following sections, we first repeat the definitions and
notations (section 1), discuss the internal consistency of slope
parameters (section 2), introduce radiative corrections and
demonstrate a method to recalculate slope parameters without reference
to raw data from older experiments (chapter 3), and finally discuss
the significance of future experiments (chapter 4).

%
\section*{Definitions and Notation}
%

In order to fix the notation, we give a short account of relevant
definitions for K$ \rightarrow 3 \pi$  decays connected to isospin
relations.

The kinematic (Dalitz plot) variables used are
$$
X = {{(s_{2}-s_{1})}\over{m_{\pi^{+}}^{2}}} \,,\quad
Y = {{(s_{3}-s_{0})}\over{m_{\pi^{+}}^{2}}}\,;
$$
$$
s_{i} = (p_{K}-p_{i})^{2} \,\,(i=1,2,3) \,,\quad
s_{0} = \frac{1}{3}(s_{1}+s_{2}+s_{3}) \,.
$$
We further use the three (not independent) variables
$r_{i}=s_{i}-s_{0}$ for expansions about the center of the Dalitz plot.
We will consider the following transition amplitudes
($K \rightarrow \pi_{1}\pi_{2}\pi_{3}$):
\[
\begin{array}{lc}
K^{+}\rightarrow \pi^{+}\pi^{+}\pi^{-}\,:\,&
A^{+}=\sum A^{+}_{IJ}(r_{1},r_{2},r_{3})\,,
\\
K^{+}\rightarrow \pi^{0}\pi^{0}\pi^{+}\,:\,&
A^{+'}=\sum A^{+'}_{IJ}(r_{1},r_{2},r_{3})\,,
\\
K^{0}_{L}\rightarrow \pi^{+}\pi^{-}\pi^{0}\,:&
A^{0}=\sum A^{0}_{IJ}(r_{1},r_{2},r_{3})\,,
\\
K^{0}_{L}\rightarrow \pi^{0}\pi^{0}\pi^{0}\,:&
A^{0'}=\sum A^{0'}_{IJ}(r_{1},r_{2},r_{3})\,.
\end{array}
\]
Factors $(n!)^{-1/2}$ for identical pions are not included. Each $A$ is
a sum of isospin contributions with $I =$ final state isospin,
contributing to the given channel, and $J = $ two times the isospin
change $\Delta I.$ The relevant states have $I = 1$ or $I = 2$ for isospin
changes
$\Delta I=\frac{1}{2},\frac{3}{2}$. The $(3\pi)_{I=0}$--state
is totally antisymmetric and therefore has a totally antisymmetric
momentum eigenfunction, that means a form factor $f\sim
(r_1-r_2)(r_1-r_3)(r_2-r_3)$, leading to contributions of at least
$3^{rd}$ order in $r_i$, which we will neglect. $I=2$ final states are
not present in
$K_{L}^{0}$ decays, if we assume CP invariance ($CP=(-1)^{I}$ for $3\pi$
 S-states).

Due to the Wigner-Eckhart-theorem, we have to introduce only 3
independent form factors $f_{11},f_{13},f_{23}$, whereby further
restrictions are derived by expanding the amplitudes up to $2^{nd}$
order and separating different symmetry classes in
$r_{1},r_{2},r_{3}$.

Defining
$f^{(1)}(r_{1},r_{2},r_{3})=\frac{1}{2}[f(r_{1},r_{2},r_{3})+f(r_{1},r_{3},r_{2})]$,
etc., the expansion up to $2^{nd}$ order in $r_{1,2,3}$ can be written:
$f^{(i)}(r_{1},r_{2},r_{3})\approx
a+br_{i}+c(r_{1}^{2}+r_{2}^{2}+r_{3}^{2})+d(2r_{i}^{2}-r_{j}^{2}-r_{k}^{2})$
(other linear and quadratic terms can be included by redefinition
of $b,c,d$, since $r_{1}+r_{2}+r_{3}=0$). After constructing properly
symmetrized $I=1,2$ final states from three pion ($I_{i}=1$) states (see
\cite {zemach} for details), one finds
\begin{eqnarray*}
A_{1J}^{+}&=&f_{1J}^{(1)}+f_{1J}^{(2)}\,,\\
A_{1J}^{0}&=&g_{1J}^{(3)}\,,             \\
A_{1J}^{+'}&=&f_{1J}^{(3)}\,,            \\
A_{1J}^{0'}&=&g_{1J}^{(1)}+g_{1J}^{(2)}+g_{1J}^{(3)}\quad (J=1,3)\,,\\
A_{23}^{+}&=&A_{23}^{+'}=2f_{23}^{(3)}-f_{23}^{(1)}-f_{23}^{(2)}
\end{eqnarray*}
with
\begin{equation}
-g_{1J}=\sqrt{2}\frac{(\frac{J}{2},+\frac{1}{2};\frac{1}{2},-\frac{1}{2}\mid
1,0)}{(\frac{J}{2},+\frac{1}{2}; \frac{1}{2},+\frac{1}{2}\mid
  1,+1)} f_{1J} \, = \, \{ {}_{-2}^{+1} \} \; f_{1J}
\end{equation}
for $J=1,3$ (the factor $\sqrt{2}$ follows from
$K_{L}^{0}=1/\sqrt{2}(1-CP)K^{0}$).

Expanding $f_{IJ}^{(i)}$ in the above manner and using
\begin{equation}
r_{3}\sim Y \,,\quad
r_{1}^{2}+r_{2}^{2}+r_{3}^{2}\sim Y^{2}+X^{2}/3 \,,\quad
2 r_{3}^{2}-r_{1}^{2}-r_{2}^{2}\sim Y^{2}-X^{2}/3\,,
\end{equation}
the four amplitudes can be written as
\begin{equation}
A=\sum [a_{IJ}+b_{IJ}Y+c_{IJ}(Y^{2}+X^{2}/3)+d_{IJ}(Y^{2}-X^{2}/3)]
\label{ampl}
\end{equation}
with the coefficients given in table 1.(we write, with slight
redefinition, again $\sum a_{IJ}\equiv a$ etc.)

The (irrelevant) relative ($-$)sign between $K^{0}$ and $K^{+}$ amplitudes
has been chosen in accordance with \cite{devlin}. We do not
distinguish at this stage between charge conjugated $K^{\pm}$ channels
and take in the following also averaged experimental data for them.

Due to strong interactions, that means rescattering between initial
and/or final states, the$f_{IJ}$ become complex functions of $r_{i}$;
therefore we have to consider relative phases between all
coefficients. Usually $a_{11}$ is chosen to be real and positive.

%
\section*{Consistency of Slope Parameters}
%

 From the usual definition of the measurable slope parameters $ g, h, k$ (in
each channel):
$$
\mid A \mid^2 \sim 1 + gY + hY^2 + kX^2
$$
one gets
\begin{eqnarray}
g = 2\, {{Re(ab^*)}\over{\mid a \mid^2}}\,,\quad
h = {{\mid b \mid^2 + 2 Re [a(c^* + d^*)]}\over{ \mid a \mid^2 }}\,,\quad
k = \frac{2}{3}\, {{Re [a(c^* - d^*)]}\over{ \mid a^2 \mid}}\,.
\nonumber
\end{eqnarray}
Since, as mentioned in the introduction, the strong phases for the
constant terms $a$ in the isospin amplitudes are thought to be small for
general reasons, the real parts of the $b$ -
coefficients (after taking $a$ to be real) are determined by the well
measured linear slopes $g$. However, it is easy to see from the isospin
relations in tab.1 that neither can the individual contributions $b_{IJ}$
be over constraint (we have 3 measured linear slope parameters $g$ for
3 quantities)
nor is it possible to get any information on their imaginary parts
or their phases, which are of primary
interest here. In order to find these, we must take into account the
quadratic slopes $h, k$. They are much more problematic from the
experimental ( statistics!) as well as from the theoretical (radiative
corrections!) point of view.\\
After defining the phases $\beta$ by
$$
b = \mid b \mid  e^{i\beta}
$$
it is easy to derive for each of the four channels a relation of the
form
\begin{equation}
h + 3k - \frac{g^2}{4 cos^2 \beta} \;=\; 4\, \frac{Re(ac^*)}{\mid a \mid^2}
\; \equiv \;R\,.
\label{rel}
\end{equation}
 From the isospin components $a_{IJ}$, $c_{IJ}$ of $a, c$ given in table 1 it
is
clear that the r.h.s. $R$ of this equation should be the same for the
two charged channels and the two neutral channels
respectively, independent of any assumptions on the phases of
$a_{IJ}$, $c_{IJ}$. It
should be approximately the same for all four channels if $c_{13} \ll c_{11}$,
according to the $\Delta I = \frac{1}{2}$ rule.

Experimental values for $g, h, k,$ and $R$, calculated from (\ref{rel})
assuming  $ \beta = 0$, are given in
table 2. Instead of using PDG values, we choose to take here in each
case the most significant (which in most cases means the latest)
 experiment. We would have to do
this anyway for the $\pi^0 \pi^0 \pi^+$ and the 3 $\pi^0$ data (which
have not yet
been included completely in the PDG tables). Moreover, because  the
consistency between some of the experimental data is certainly
questionable (compare \cite{devlin} and \cite{pdg}), the errors may be more
consistent; furthermore it becomes possible to compare our estimate for the
effect of the Coulomb corrections with that given in the original
paper \cite{ford}.
In any case, the PDG values do not disagree significantly from those
used here.

While the $R$ -- values from the two $K^0_L$ decay experiments are in perfect
agreement, the situation for the charged $K$ -- decays is less
clear, due to the lower (by one order of magnitude) statistics in the
$\pi^0\pi^0\pi^+$
channel. If
we assume $c_{13}\ll c_{11}$ and take the $R$ value from the (most
significant) $\pi^0\pi^0\pi^0$ -- experiment \cite{somalvar}, we find
linear relationships between $h$ and $k$ for all the other channels, as
shown in fig.1. The measured $(h, k)$ point should fall above the
line, if $\beta\neq 0$.
The consistency between the $\pi^{\pm} \pi^{\pm} \pi^{\mp}$ and the
$\pi^0 \pi^0 \pi^0$ data is rather bad; one has to assume a fluctuation of
$\sim$ 2 st.dev., or a ratio of $c_{13}$ / $c_{11}$ = $0.3 \pm 0.1$,
in order to reach agreement. However from a comparison of figs.1a,b
which show the results from the same experiment derived without and with
Coulomb corrections, respectively,
it is clear that this result is very sensitive to these corrections.
Therefore we will introduce more complete radiative
corrections in the next section.

%
\section*{Radiative Corrections}
%

Radiative corrections for hadronic processes are, generally speaking,
model dependent, insofar as structural effects are concerned. The
leading contributions
can be estimated, however, by calculating the virtual photon exchanges
and soft photon emissions by point--like mesons, describing the weak
decay by the most simple local interaction:
\begin{equation}
L_w = c_w\,K \, \pi_1 \,\pi_2\, \pi_3 \,.
\nonumber
\end{equation}
To first order in $\alpha$, the relevant graphs are shown in fig.3a,
while structural contributions fig.3b are not considered.

 All analytic expressions needed are given below for easy
reference. A FORTRAN code for calculating the correction factor
as a function of $X$ and $Y$  for all $K \rightarrow 3\pi$ channels
is available on request (schaale@ifh.de).

The corrected decay probability is given by
\begin{equation}
d \Gamma(s_i) = d \Gamma_0(s_i) \bigg\{1 + {{\alpha}\over{2\pi}}
\bigg[\sum_{i=1}^3 e_0 e_i F_{K\pi_i} -
     \sum_{i<j} e_i e_j F_{\pi_i\pi_j} +
     \sum_{i=1}^3 e_i^2 F_i
\bigg]\bigg\}\; ,
\label{rad}
\end{equation}
where $e_i = 0,\pm 1$ are charge factors of $K,\pi_i$;
\begin{eqnarray}
F_{K\pi_i} =&-&ln\bigg({{2 \Delta\varepsilon}\over{\mu}}\bigg)^2
\bigg(2 + {{1}\over{v_i}}\; ln {{1-v_i}\over{1+v_i}}\bigg)
+ \bigg(1-{{\Sigma}\over{s_i}}\bigg)\bigg[{{v_i}\over{2}}
\;ln\bigg({{1-v_i}\over{1+v_i}}\bigg)
\nonumber
\\
&+&2\int_{-1}^{+1} {{dz}\over{\Phi_i(z)}}\;\bigg(\;ln
\bigg|{{s_i}\over{4\mu^2}}
\Phi_i(z)\bigg| - Q_i(z)\;
ln\bigg|{{1-Q_i(z)}\over{1+Q_i(z)}}\bigg|\bigg)\bigg]
\nonumber
\\
&-&\bigg(1 - \frac{\Delta}{s_i}\bigg)\;ln{{m}\over{\mu}}
 - 6 {{\mu}\over{m-\mu}}\;ln {{m}\over{\mu}} + 8
\nonumber
\end{eqnarray}
and
\begin{eqnarray}
F_{\pi_i\pi_j} =&-&ln\bigg({{2 \Delta\varepsilon}\over{\mu}}\bigg)^2
\bigg(2 + {{1+v_{ij}^2}\over{v_{ij}}}\; ln {{1-v_{ij}}\over{1+v_{ij}}}\bigg)
- \bigg(v_{ij}\; ln {{1-v_{ij}}\over{1+v_{ij}}} + 2\bigg)
\nonumber
\\
&+&\big(1+v_{ij}^2\big)\int_{-1}^{+1} {{dz}\over{z^2-v_{ij}^2}}
\bigg(\;ln \bigg|{{z^2-v^2_{ij}}\over{1-v^2_{ij}}}\bigg|
- {{z^2-v^2_{ij}}\over{z^2v^2_{ij}-1}}
 Q_{ij}(z)\;ln \bigg|{{1-Q_{ij}(z)}\over{1+Q_{ij}(z)}}\bigg|\bigg) -
 8{\bf G}\; ,
\nonumber
\end{eqnarray}
where ${\bf G} = 0.915966$ is Catalan's constant;
$$
F_i = - \frac{1}{v_i}\;ln{{1-v_i}\over{1+v_i}} \; - 2 \; ;
$$
$$
\Phi_i(z) = z^2 - 2 z \frac{\Delta}{s_i} - \bigg(1 -{{2\Sigma}\over{s_i}}\bigg)
\,,\quad \Sigma = m^2 + \mu^2 \,,\quad \Delta = m^2 - \mu^2 \; ;
$$
$$
Q_i(z) = {{(1+z)\varepsilon_i + (1-z)m}\over{(1 + z)|\vec{p}_i|}}\,,\quad
Q_{ij}(z) = {{(1+z)\varepsilon_i + (1-z)\varepsilon_j}
\over{|(1 + z)\vec{p}_i + (1-z)\vec{p}_j|}}\,;
$$
$v_i = |\vec{p}_i|/\varepsilon_i$ is the CMS-velocity;
$v_{ij} = \sqrt{1 -4\mu^2/s_k}$ are the velocities in
$(i,j)$-rest system; $m = m_{K^+}$, $\mu =m_{\pi^+}$, and
$\Delta \varepsilon$ is the $\gamma$  cut--off energy.
The conventional Coulomb correction factor \cite{coul} is given by
\begin{equation}
\prod_{i<j}\eta_{ij}/(exp(\eta_{ij})-1)\,,\quad
\eta_{ij}=2\pi\alpha e_ie_j/|\vec{v}_i-\vec{v}_j|\,.
\label{coul}
\end{equation}

In fig.4 are shown the correction factors according to (\ref{rad}), with
$\Delta \varepsilon = 10 MeV$ are shown as functions of $X,Y$ for all
charged channels. They are largest for the  $K^\pm \to
\pi^{\pm}\pi^{+}\pi^{-}$ decay (for this channel we show for
comparison also the
values calculated with $\Delta \varepsilon = 50 MeV$ and with the
conventional Coulomb factor (\ref{coul})).
For the other decay modes the corrections are much less important and
depend only on $Y$. For  $K^\pm \to \pi^{\pm}\pi^{0}\pi^{0}$ the
Coulomb factor (\ref{coul}) equals $1$, because there is only one
charged particle in the final state.

We add some remarks concerning the well known
singularities appearing in the treatment of radiative corrections, in
order to show the limits of this approach.

The first kind are ultraviolet divergences of the loop integrals which
result in explicit cut--off dependent terms
 $\sim\,ln{(\Lambda/\mu)}$.
The reason is that we have chosen  purely local effective
interactions without hadronic form factors, which would regularize
these integrals. In our approach we renormalized them
by the requirement that their contributions should disappear for
transitions, in which the incoming charge reemerges as outgoing with the
same velocity. This means we subtract
$F_{K\pi}\big(s_i=(m-\mu)^2\big)$ and $F_{\pi\pi}(s_i=0)$ respectively
(standard renormalization on mass shell)\footnote{The subtraction
  constants include also all contributions from photon emission
  graphs; in this respect the present results differ from those given
  in \cite{old paper}}.
As we neglected
higher order corrections to $L_w$ ($L=L_w + \eta L_1$, with $\eta \sim
0.1$, say),
being responsible for the kinematic structure of the Dalitz--plot,
our results for slope parameters are correct up to terms of order
$\alpha\eta$. The constant terms $a_{IJ}$ are affected by the unknown
renormalization ambiguity of order $\alpha$; in other
words, hadron structure effects may differ between $K^0_L\to3\pi^0$ and
 charged K decays by terms of order
$\alpha$. It is easy to show that this induces also corrections of
order $\alpha\eta$ for slope parameters. Their treatment would require a more
complicated effective Lagrangian, inclusion of structural photons etc.,
and would be strongly model--dependent.

A second kind are so--called collinear singularities, appearing in the case of
at least two charged particles with equal velocities in the final
state, that means on the Dalitz--plot boundary. In these
points the perturbation expansion breaks down.
Accepting that some regions of the
Dalitz-plot are just not handled correctly by the theory, some caution
in the treatment of experimental data near these singularities may be
required. It can be shown that no infinities are
encountered in integrals over kinematic regions (integrability).

The third
and last type are infrared divergences caused by low energy photons.
Their cure is well known, leading to the introduction of the upper
limit $\Delta\varepsilon$, defined here in the K rest system, up to
which weak photons are to be included in the definition of the
decay channel. A rough estimate of $\Delta\varepsilon$ may be derived
from the mass resolution achieved in reconstructing the K mass.
However, care should be exercised when high energy
K--decays are analyzed. A correct experimental treatment would have to include
radiative processes in the Monte--Carlo and to establish an
effective $\Delta\varepsilon$ in this way.
\footnote{There are significant differences, at least concerning
  the parameter $k$, between \cite{ford} and a later
  experiment \cite{devaux},
which found $k = -0 .0205 \pm 0.0039$ to be compared with $-
  0.0075 \pm 0.0019$ \cite{ford} for $K^+ \to \pi^+\pi^+\pi^-$. As the
  first experiment measured only the momentum of the odd pion in the
  final state, the second one all momenta, effects of the above mentioned
  kind may be present. \cite{devaux} do not present data without
  (Coulomb) correction, therefore we used only \cite{ford}.}

%

In order to estimate the influence of the radiative and the Coulomb corrections
on the Dalitz--plot parameters without reference to the
experimental data, i.e. uncorrected Dalitz--plot densities, we
introduce moments $<X^m \cdot Y^n>^{(')}$ with respect to normalized
Dalitz--plot
densities $p(X,Y)$ and $p'(X,Y)$, where $p$ represents a constant
density and $p'$ includes radiative corrections to this constant
density. After expanding $p'$ in $X,Y$, it is easy to get approximate
expressions for $p'$ moments $<...>'$ in terms of $p$ moments
$<...>$, but for our case we have calculated all relevant moments
numerically. If we suppose that the experimental data sample is
already corrected for experimental efficiency and
background
\footnote{In the actual evaluation of an experiment, $p$ would be
  taken from a
Monte--Carlo simulation of the measured Dalitz--plot distribution,
using a constant Dalitz--plot density as input.}.
, the slope parameters are to be derived from a fit to the
Dalitz--plot density, which  may be written as
\begin{equation}
f(X,Y) = {{1+\vec{v}\cdot\vec{g}}\over{1+<\vec{v}> \cdot\vec{g}}} \; p(X,Y)\,,
\nonumber
\end{equation}
where for convenience the vectors
\[
\vec{v} =
\left(
\begin{array}{c}
   Y
\\ Y^2
\\ X^2
\end{array}
\right)
\hspace{2cm} \vec{g} =
\left(
\begin{array}{c}
g \\ h \\ k
\end{array}
\right)
\]
are introduced. We write
$\vec{v}(i) = \vec{v}(X_i,Y_i)$ for measured $X_i$ and $Y_i$ for the
$\mbox{i}^{th}$ event.
$\vec{g}$ is to be estimated by the Maximum Likelihood method from
the Likelihood function
\begin{equation}
L = \prod_{i=1}^{n} f(X_i,Y_i)
\nonumber
\end{equation}
leading to the system of equations for $\vec{g}$
\begin{equation}
\frac{1}{n} \sum_i
{{\vec{v}(i)}\over{1+\vec{v}(i)\cdot\vec{g}}} =
{{<\vec{v}>}\over{1+<\vec{v}> \cdot\vec{g}}}
\nonumber
\end{equation}
(we do not distinguish here between estimates and population values).
If we now apply a further correction, e.g. $p \rightarrow p'$, we get,
with the same sample of experimental data $X_i,Y_i$, corrected
parameters $g',h',k'$, which can be
expressed  in terms of $\vec{g}$ and moments $<...>, <...>'$.
The equations for $\vec{g}'$ are:
\begin{equation}
\frac{1}{n} \sum_i
{{\vec{v}(i)}\over{1+\vec{v}(i)\cdot \vec{g}'}} =
{{<\vec{v}>'}\over{1+<\vec{v}>'\cdot \vec{g}'}}\,.
\nonumber
\end{equation}
For small corrections we may expand both sides in
$\Delta \vec{g} = \vec{g}' -\vec{g}$ and obtain a linearized set of equations:
\begin{equation}
V \cdot \Delta \vec{g} =
{{<\vec{v}>}\over{1+<\vec{v}>\cdot \vec{g}}} -
{{<\vec{v}>'}\over{1+<\vec{v}>'\cdot \vec{g}}}\,,
\label{equ}
\end{equation}
where the symmetric matrix $V$ has the elements
\begin{equation}
V_{kl}=
\frac{1}{n} \sum_i \Big(
{ { v_k v_l }\over{ (1+\vec{v}\cdot \vec{g})^2 } }
\Big)_{\vec{v}=\vec{v}(i)} -
{{<v_k>' <v_l>'}\over{(1+<\vec{v}>'\cdot \vec{g})^2}} \quad (k,l=1,2,3)\,.
\nonumber
\end{equation}
The dependence on $X_i, Y_i$ can now be eliminated by replacing
$\frac{1}{n} \sum$ with $\int dX dY f(X,Y)$, leading to
\begin{equation}
V_{kl}=
{{1}\over{1+<\vec{v}>\cdot \vec{g}}}
\Big<
{{v_k v_l}\over{1+\vec{v}\cdot \vec{g}}}
\Big>
 - {{<v_k>' <v_l>'}\over{(1+<\vec{v}>'\cdot \vec{g})^2}}\,,
\nonumber
\end{equation}
where the first terms on the right--hand side are also to be evaluated
for given $g,h,k$ numerically.

Surely one has to be aware of the severe limitations of the above
approach if, for experimental or other reasons, the uncorrected
parameters $\vec{g}$ do not represent the density over the whole
Dalitz--plot. To give an extreme example, suppose they had been
derived from a fit to the Dalitz--plot density in a region where the
corrections disappear. They were then found identical to the
corrected parameters, if the "corrections" are applied to the raw data
sample. For our method one has to assume however that the uncorrected
parameters fit the density equally well over the whole Dalitz--plot.
If, as we may further suppose in our example, there are sizeable
corrections to the density in the unmeasured region this is not the
case, and consequently we find the corrected parameters different
from the uncorrected ones, possibly even outside the statistical
errors.

As a check of our method we calculated for the conventional
Coulomb correction factor the corrected parameters $\vec{g}'$,
corresponding to the first column of table 2, from the uncorrected
values of ref.\cite{ford} in the second column. The results are given in
table 3 together with the differences with respect to the  corrected
values from ref.\cite{ford}.

Our conclusion from this comparison is that the method is useful to
demonstrate the influence of radiative corrections on the quadratic slope
parameters, where statistical errors are relatively large. It is not a
substitute, however, for a complete (re)analysis of precision experimental data
like those existing for linear slope
parameters, for which systematic corrections are more subtle.

%
\section*{Results and Conclusions}
%

The results for $\Delta g, \Delta h, \Delta k$ are shown as
a function of the linear slope $g$ in fig.5
and for the actual $g$--values (see table
2) in table 4. The dependence on $h$ and $k$ is for small values of these
parameters negligible, they are set to zero everywhere.

For the $K^\pm \to \pi^{\pm}\pi^{+}\pi^{-}$ channel the discrepancy
of the $R$--value with that for $K^{0}_{L} \to 3\pi^0$ (see table 2)
disappears after applying the radiative corrections (\ref{rad}) instead of
the Coulomb--factor (\ref{coul})(compare fig.1a) with fig.2). We find
$$
R^{\pm +-}_{rad.corr.}(\beta =0)= -.0078 \pm .0089 ,
$$
indicating that the quadratic coefficients $c$ in the decay amplitude
(\ref{ampl}) may be small and of the same order as for $K^0_{L}$--decays.

The corresponding corrections for the channel $K^\pm \to
\pi^{\pm}\pi^{0}\pi^{0}$ are smaller than the numerical accuracies of
the numbers in table 2. A comparison of $R$--values of the different
charged K--decay channels with comparable statistics
would be interesting.

In order to derive some first information on
possible phases $\beta$ we may use the relation
$$
R(\beta) = R(0) - \frac{g^2}{4}\tan^2\beta \,.
$$
For the  $K^{0}_{L} \to \pi^{+}\pi^{-}\pi^{0}$--channel we can
identify $R^{+-0}(\beta)$ with $R^{000}$ and find from the last row of
tab. 2\footnote{radiative corrections by the authors \cite{messner} are
already included}
$$
\tan^2\beta = 0.0 \pm .086 \, ,
$$
i.e. a phase angle
$\beta^{+-0}\leq 16^o$. For the charged kaon decay $K^\pm \to
\pi^{\pm}\pi^{+}\pi^{-}$, assuming $c_{13}\ll c_{11}$ and comparing
$R^{000}$ also with $R^{\pm
  +-}_{rad.corr.}$, we find analogously $\beta^{\pm +-}\leq 43^o$.

Despite comparable errors of the $R$--values for the two cases, the
restriction on
$\beta^{\pm +-}$ is weaker than that found on $\beta^{+-0}$. This is
due to the different linear slopes $g$. In view of this, and also because
it is much less influenced by radiative corrections, the channel
$K^\pm \to \pi^{\pm}\pi^{0}\pi^{0}$ (with $g \sim 0.6$) deserves
special attention by experiment. Enhancing the data sample for this
channel by an order
of magnitude (to $\sim 5\cdot 10^5$ events) could lead to a
determination of the phase $\beta$ with an error $\leq 15^o$ similar as for
$K^{0}_{L}$ decays. Clearly, this would help a lot to constrain
effective Lagrangian models with regard to higher order
($p^4$--, loop--, penguin--) contributions, especially if taken together
with results from radiative K--decays, where also new
experimental and theoretical work is going on.

Besides this, one should be aware that also for the other channels
considered here one has to rely presently on only few large statistics
experiments. Concerning the presentation of new data, we would like to
advocate to publish the data also in a form uncorrected for radiative or
Coulomb effects. Furthermore, one should clearly state the regions of
the Dalitz--plot to which the slope parameters have been fitted. There is room
for later improvements of the radiative corrections, including structural
radiation, after more realistic effective Lagrangians will have been
introduced.

    One of the authors (A.A.Bel'kov) acknowledges the support from
DFG, Project Eb 139/1--1.

\newpage
\begin{table}[p]\centering
\begin{tabular}{|c|c|c|c|c|}
\hline
Channel&$+ + -$&$0 0 +$&$+ - 0$&$0 0 0$\\
\hline
$a$&$2(a_{11}+a_{13})$&$a_{11}+a_{13}$&$-(a_{11}-2a_{13})$&$-3(a_{11}-2a_{13})$\\
$b$&$-(b_{11}+b_{13})+b_{23}$&$b_{11}+b_{13}+b_{23}$&$-(b_{11}-2b_{13})$&$0$\\
$c$&$2(c_{11}+c_{13})$&$c_{11}+c_{13}$&$-(c_{11}-2c_{13})$&$-3(c_{11}-2c_{13})$\\
$d$&$-(d_{11}+d_{13})+d_{23}$&$d_{11}+d_{13}+d_{23}$&$-(d_{11}-2d_{13})$&$0$\\
\hline
\end{tabular}
\caption{Isospin Amplitudes}
\end{table}
\begin{table}[p]\centering
\begin{tabular}{|c|c|c|c|c|c|} \hline
Ch.&\multicolumn{2}{|c|}{$\pi^\pm \pi^\pm \pi^\mp$}&
$\pi^0\pi^0\pi^+$&$\pi^+\pi^-\pi^0$&
$\pi^0\pi^0\pi^0$ \\
\hline
Expt.&\cite{ford},a&\cite{ford},b&\cite{bolotov}&\cite{messner}&\cite{somalvar}\\ \hline
$g$ & $-.2173\pm.0026$&$-.1866\pm .0025$&$.575\pm.022$&$.677\pm.010$&0 \\
$h$ & $.0156\pm.0062$&$.00125\pm .0062$&$.021\pm
.023$&$.079\pm .007$&$-.0033\pm .0013$ \\
$k$ & $-.0079\pm .0019$&$.0029\pm .0021$&$.011\pm
.007$&$.0097\pm .0018$&$h/3$ \\ \hline
$R$ & $-.0199\pm .0084$&$.0013\pm .0088$&$-.029\pm
.032$&$-.0065\pm .0095$&$-.0066\pm.0026$ \\ \hline
\end{tabular}
\caption{Experimental Data (a with, b without Coulomb correction)}
\end{table}
\begin{table}[p]\centering
\begin{tabular}{l|c|c|} \cline{2-3}
   &calc. by (\ref{equ})&Diff. to \cite{ford}\\ \hline
\vline$\;\;g$&$ -.2236\pm .0025$&$.0063\pm .0037$\\
\vline$\;\;h$&$.0149\pm.0062$&$.0007\pm .0088$\\
\vline$\;\;k$&$-.0079\pm .0021$&$0.0\pm .003$\\ \hline
\end{tabular}
\caption{Comparison for Coulomb corrections}
\end{table}
\begin{table}[p]\centering
\begin{tabular}{|c|c|c|c|} \hline
Channel&$\pm + - $&$\pm 0 0$&$+ - 0$\\ \hline
$\Delta g$&$-1.99\cdot 10^{-2}$&$-1.45\cdot 10^{-3}$&$7.02\cdot
10^{-3}$\\
$\Delta h$&$7.28\cdot 10^{-3}$&$-8.47\cdot 10^{-4}$&$1.63\cdot
10^{-3}$\\
$\Delta k$&$-4.72\cdot 10^{-3}$&$7.66\cdot 10^{-7}$&$-1.70\cdot
10^{-5}$\\
\hline
\end{tabular}
\caption{Radiative Corrections for Slope Parameters}
\end{table}

\newpage
  %
   
   %
\newpage
\noindent
{\Large \bf Figure captions}
\begin{enumerate}
\item[{\bf Fig. 1}]
Plots of quadratic slope parameters $k$ vs. $h$ from table 2. The linear
relation (\ref{rel}) is indicated with 1 s.d. errors. \\
   \begin{tabular}{rp{10cm}}
$++-$ a&:  channel $K^{\pm} \to \pi^{\pm} \pi^+ \pi^-$ with
     Coulomb corrections (\ref{coul})\\
      b&: uncorrected \\
$+\;0\;0$&: channel $K^{\pm} \to \pi^{\pm} \pi^0 \pi^0$, uncorrected\\
$+\;-\;0$&: channel $K^0_L \to \pi^+ \pi^- \pi^0$, corrected by
\cite{messner}
   \end{tabular}
\item[{\bf Fig. 2}]
 The same as fig.1 for channel $K^{\pm} \to \pi^{\pm} \pi^+ \pi^-$
 corrected according to (\ref{equ}) with rad. corr. (\ref{rad})\\
\item[{\bf Fig. 3}]a) Graphs for radiative corrections to first order\\
b) Graphs with inclusion of structural radiation
\item[{\bf Fig. 4}] Plots of correction factors as functions of $X,Y$\\
a) Rad. corr. for  $K^{\pm} \to \pi^{\pm} \pi^+ \pi^-\;,\Delta
\varepsilon=10MeV,50MeV$ (broken lines)\\
b) Coul. corr. for  $K^{\pm} \to \pi^{\pm} \pi^+ \pi^-$\\
c) Rad. corr. for $K^{0}_{L} \to \pi^{+} \pi^- \pi^0$\\
d) Rad. corr. for $K^{\pm} \to \pi^{\pm} \pi^0 \pi^0$
\item[{\bf Fig. 5}]
a) - c) Dependence of $\Delta g, \Delta h, \Delta k$ on $g$ (for $h = k
= 0 $) for the channel $K^{\pm} \to \pi^{\pm} \pi^+ \pi^-$
\end{enumerate}
\end{document}